# Molecular theory of graphene

E. F. Sheka

Peoples' Friendship University of Russia, Miklukho-Maklay, 6, Moscow 117198, Russia

sheka@icp.ac.ru

Abstract:   Odd electrons of benzenoid units and correlation of these electrons having different spins are the main concepts of the molecular theory of graphene. In contrast to the theory of aromaticity, the molecular theory is based on the fact that odd electrons with different spins occupy different places in the space so that the configuration interaction becomes the central point of the theory. Consequently, a multi-determinant presentation of the wave function of the system of weakly interacting odd electrons is absolutely mandatory on the way of the theory realization at the computational level. However, the efficacy of the available CI computational techniques is quite restricted in regards large polyatomic systems, which does not allow performing extensive computational experiments. Facing the problem, computationists have addressed to standard single-determinant ones albeit not often being aware of how correct are the obtained results. The current chapter presents the molecular theory of graphene in terms of single-determinant computational schemes and discloses how reliable information about electron-correlated system can be obtained by using either UHF or UDFT computational schemes.

When the paper was written, a beautiful conceptually profound 'informal reflection' of Roald Hoffmann appeared in the first issue of the Angewandte Chemie (International Eddition) that celebrates its 125-year anniversary [1]. Hoffmann's "Small but Strong Lessons from Chemistry for Nanoscience" turned out to be remarkably concordant to the main ideas discussed in the current paper. This should be expected since the Hofmann concepts on stabilizing singlet states of biradicals in organic chemistry (see [2] and references therein) and dimeric molecular magnets [3] have led the foundation of the molecular theory of fullerenes [4, 5], application of which to graphene science is discussed below. These problems are on knife-edge today that is why, once fully agree with Hoffmann's answers to the question 'What you can trust about theory?', I would like to preface the presentation of the text of a quote from the 'informal reflection', placing it as the epigraph

> It goes without saying that theory is really of value when it is used to perform numerical experiments that capture a trend. Not numbers, but a trend.
>
> Roald Hoffmann, 2013.

## Introduction

For more then ten years I have been immersed in an absorbing world of quantum chemistry of $sp^2$ nanocarbons, the world full of mysteries, hidden obstacles, and wonderful discoveries. My first travelling was stimulated by a wish to find answer to a very simple question: why there is no fullerene $Si_{60}$ while fullerene $C_{60}$ does exist? A widely spread standard statement that "silicon does not like $sp^2$ configuration" just postulated the fact but did not explain the reason of such behavior. Moreover, computations, available by that time, showed that $Si_{60}$ molecule could exist. A comparative examination of $C_{60}$ and $Si_{60}$ showed a strange feature in the high-spin states behavior of the molecules. As occurred, a sequence of spin-varying states, singlet (RHF)-triplet-quintet formed a progressively growing series by energy for the $C_{60}$ molecule while for the $Si_{60}$ one energy of the triplet and quintet states turned out to drop drastically with respect to the RHF singlet. Due to a crucial controversy with the reality, a natural question arose: what is wrong with the molecule singlet state? I will not touch here a frequent claiming that semiempirical approach is bad. It is not the case, in general, and is absolutely non relevant to carboneous and siliceous species due to superior parameterization of both atoms. Actually, all the next stories have shown that the matter was not in the wrong approximation but was provided by an inherent peculiarity of both molecules. At that time, in 2003 was shown that the singlet state of the $Si_{60}$ molecule took its correct place below the triplet one if only it is calculated by using open-shell unrestricted Hartree-Fock (UHF) approximation [4, 6, 7]. Since then in more than eight decades of papers that follows, I and my colleagues have been convinced ourselves and have tried to convince others that UHF approach touches very intimate properties of $sp^2$ nanocarbons that select them from other carboneous species and put them in a particular place. The properties are result of a significant weakening of interaction between odd electrons of the species in comparison with, say, that one in the benzene molecule.

During these investigations was obtained answer to the initial question concerning the absence of $Si_{60}$ molecule [4], were disclosed regulations that govern chemistry, magnetism, biomedical and photonic behavior of carboneous fullerenes [5], was shown a tight similarity in the description of the properties of fullerenes, carbon nanotubes, and graphene molecules [5, 8]. Little by little an applied molecular theory of $sp^2$ nanocarbons becomes sharply defined, which has revealed itself the most vividly in case of graphene. However, graphene, which is a famous nobeliated 2D solid, and molecular theory- if there is no controversy between these subjects?

The answer lies on a surface and follows from a well known Wiki definition of graphene: 'Graphene is an allotrope of carbon, whose structure is one-atom-thick planar sheets of $sp^2$-bonded carbon atoms that are densely packed in a honeycomb crystal lattice [9]'. This definition clearly exhibits a molecular-crystal duality of this extraordinary substance. From the molecular viewpoint, the extraordinariness is provided with the availability of odd electrons that are responsible for the $sp^2$ configuration of valence electrons of carbon atoms. 2D-dimensionality, on the other hand, dictates peculiar properties of a regularly packed honeycomb pattern. Due to this, the graphene properties are similar to those of both polycondensated benzenoid molecules and 2D-dimensional crystals. Obviously, fundamental characteristics of the two forms are tightly interconnected. Thus, as will be discussed below, such seemingly solid state properties as magnetism and mechanics of graphene are of molecular origin.

This peculiar duality is embodied in the computational strategy of graphene, as well. On one hand, the solid state microscopic theory of quasiparticles in 2D space forms the ground for the description of graphene crystal. On the other hand, quantum molecular theory creates the image of the graphene molecule. Seemingly, the two conceptual approaches, obviously different from the computational viewpoint, have nevertheless much in common. Thus, the solid state quasiparticles are usually described in the approach based on a unit cell and/or supercell followed with periodic boundary conditions; besides, the unit cell is described at the molecular theory level thus presenting a molecular object similarly to the case of the molecular theory. However, the very molecular object provides a crucial difference between the two approaches. In the case of a correct solid state approach, the cell and/or supercell should be strictly chosen as a known crystalline motive. Accordingly, two-atomic cell of graphene crystal finds its exhibition in the peculiarities of electron band structure of the crystal. However, nowadays, the solid state approach is explored in the graphene science in regards practically all the phenomena including graphene chemical modification, graphene deformation and magnetization. The two-atomic unit cell of the crystal does not meet conditions needed for examining these complicated events, particularly, related to chemical modification. The cell is substituted by a supercell, whose structure is taken voluntary, being in the majority of the case just 'drawn' in stead off attributed to a reality. Moreover, regular structure of the graphene object is fastened by periodical boundary conditions. The two features of the solid-state approach, namely, a voluntary chosen supercell and the fastened periodicity make clear the Hoffmann answer "Not much" to the question "What you can trust about theory?" [1]. Then Hoffmann continues: "Aside from the natural prejudice for simplicity, people really want translational periodicity in their calculations, for then the quantum mechanical problem reduces to one of the size of the unit cell. But the real world refuses to abide by our prejudices. And it is often an aperiodic, maximally defect-ridden, amorphous world, where emergent function is found in matter that it is as far from periodic as possible". The reality of graphene science, particularly, related to chemical modification, strongly witnesses the domination of aperiodic structures. In view of this, the molecular theory of graphene has a convincing preference since its molecular object is created in due course of computations without structural restrictions introduced in advance.

The current paper is concentrated at the molecular essence of graphene considered from the viewpoint of the molecular theory of $sp^2$ nanocarbons. The theory is based on two main concepts, which involve the odd-electron origin of the graphene electron system and these electrons correlation. The latter turns out to play the governing role. As will be shown below, such an approach occurs very efficient in describing chemical, magnetic, mechanical, and optical properties of graphene.

## Odd electrons correlation

In spite of formally two-atomic unit cell of crystalline graphene, its properties are evidently governed by the behaviour of odd electrons of hexagonal benzenoid units. The only thing that we know about the behaviour for sure is that the interaction between odd electrons is weak; nevertheless, how weak is it? Is it enough to provide a tight covalent pairing when two electrons with different spins occupy the same place in space or, oppositely, is it too weak for this and the two electrons are located in different spaces thus becoming spin correlated? This supremely influential molecular aspect of graphene can be visualised on the platform of molecular quantum theory.

To exhibit a trend, a system computational experiment must be carried out meaning that a vast number of computations are to be performed as well as a great number of atoms are to be considered. When speaking about electron correlation, one should address the problem to the configuration interaction (CI). However, neither full CI nor any its truncated version, clear and transparent conceptually, can be applied for computational experiments, valuable for graphene nanoscience. Owing to this, techniques based on single unrestricted open-shell determinants becomes the only alternative. Unrestricted Hartree-Fock (UHF) and unrestricted DFT (spin polarized, UDFT) approaches form the techniques ground and are both sensitive to the electron correlation, but differently due to different dependence of their algorithms on electron spins [10, 11]. Application of the approaches raises two questions: 1) what are criteria that show the electron correlation in the studied system and 2) how much are the solutions of single-determinant approaches informative for a system of correlated electrons.

Answering the first question, three criteria, which highlight the electron correlation at the single-determinant level of theory, can be suggested. Those concern the following characteristic parameters:

**Criterion 1:**

$$\Delta E^{RU} \geq 0,$$

where

$$\Delta E^{RU} = E^R - E^U \quad (1)$$

presents a misalignment of energy. Here, $E^R$ and $E^U$ are total energies calculated by using restricted and unrestricted versions of the program in use.

**Criterion 2:**

$$N_D \neq 0,$$

where $N_D$ is the total number of effectively unpaired electrons and is determined as

$$N_D = tr D(r|r') \neq 0 \quad \text{and} \quad N_D = \sum_A D_A. \quad (2)$$

Here, $D(r|r')$ [12] and $D_A$ [13] present the total and atom-fractioned spin density caused by the spin asymmetry due to the location of electrons with different spins in different spaces.

**Criterion 3:**

$$\Delta \hat{S}^2 \geq 0, \qquad (3)$$

where

$$\Delta \hat{S}^2 = \hat{S}_U^2 - S(S+1)$$

presents the misalignment of squared spin. Here, $\hat{S}_U^2$ is the squared spin calculated within the applied unrestricted technique while $S(S+1)$ presents the exact value of $\hat{S}^2$.

Criterion 1 follows from a well known fact that the electron correlation, if available, lowers the total energy [14]. Criterion 2 highlights the fact that the electron correlation is accompanied with the appearance of effectively unpaired electrons that provide the molecule radicalization [12, 13, 15]. Those electrons total number depends on interatomic distance: when the latter is under a critical value $R_{cov}^{crit}$, two adjacent electrons are covalently bound and $N_D = 0$. However, when the distance exceeds $R_{cov}^{crit}$, the two electrons become unpaired, $N_D \geq 0$, the more, the larger is the interatomic spacing. In the case of the $sp^2$ C-C bonds, $R_{cov}^{crit}$ =1.395Å [16]. Criterion 3 is the manifestation of the spin contamination of unrestricted single-determinant solutions [13, 15]; the stronger electron correlation, the bigger spin contamination of the studied spin state.

Table 1 presents sets of the three parameters evaluated for a number of graphene molecules presented by rectangular $(n_a,n_z)$ fragments of graphene ($n_a$ and $n_z$ count the benzenoid units along armchair and zigzag edges of the fragment, respectively [19]), $(n_a,n_z)$ nanographenes (NGrs) below, by using AM1 version of semiempirical UHF approach implemented in the CLUSTER-Z1 codes [18]. To our knowledge, only these codes allow for computing all the above three parameters simultaneously. As seen in the table, the parameters are certainly not zero, obviously greatly depending on the fragment size while their relative values are practically non size-depending. The attention should be called to rather large $N_D$ values, both absolute and relative. The finding evidences that the length of C-C bonds in the considered molecules exceed the critical value $R_{cov}^{crit}$ =1.395Å. It should be added as well that the relation $N_D = 2\Delta \hat{S}^2_U$, which is characteristic for spin contaminated solutions in the singlet state [13], is rigidly kept over all the fragments.

**Table 1.** Identifying parameters of the odd electron correlation in rectangular graphene fragments [17]

| Fragment $(n_a, n_z)$ | Odd electrons $N_{odd}$ | $\Delta E^{RU}$ * kcal/mol | $\delta E^{RU}$ % ** | $N_D, e^-$ | $\delta N_D, \%$ ** | $\Delta \hat{S}^2_U$ |
|---|---|---|---|---|---|---|
| (5, 5)   | 88  | 307  | 17 | 31    | 35 | 15.5  |
| (7, 7)   | 150 | 376  | 15 | 52.6  | 35 | 26.3  |
| (9, 9)   | 228 | 641  | 19 | 76.2  | 35 | 38.1  |
| (11, 10) | 296 | 760  | 19 | 94.5  | 32 | 47.24 |
| (11, 12) | 346 | 901  | 20 | 107.4 | 31 | 53.7  |
| (15, 12) | 456 | 1038 | 19 | 139   | 31 | 69.5  |

* AM1 version of UHF codes of CLUSTER-Z1 [18]. Presented energy values are rounded off to an integer

* The percentage values are related to $\delta E^{RU} = \Delta E^{RU} / E^R(0)$ and $\delta N_D = N_D / N_{odd}$, respectively

Summarizing said above it is possible to conclude the following.

1. Nowadays, single-determinant computational schemes, based on open-shell approximation of either Hartree-Fock or DFT approach, are the only alternative for practically valuable computations of polyatomic graphene systems (Nat>30-40);
2. For electron-correlated systems, the obtained solutions are not exact but spin-mixed;
3. The question arises: which reliable information about electron-correlated system can be obtained by using either UHF or UDFT computational scheme?

Given below has been organized as getting answers to this question.

## *Answer 1. Broken symmetry approach allows obtaining the exact energy values of pure-spin states*

The wave functions of the unrestricted single-determinant solutions satisfy the operator equations for energy and *z*-projection of spin $S_z$ but do not satisfy the operator equation for squared spin $\hat{S}^2$. This causes a spin contamination of the solution whose extent is determined by $\Delta \hat{S}^2$ (3). Owing to this, one faces the problem of the evaluation of the energies of pure spin states.

The unrestricted broken symmetry (UBS) approach suggested by Noodleman [20] can be considered as the best way to solve the problem. It is the most widely known among the unrestricted single-determinant computational schemes used in practice, both UHF and UDFT. The UBS approach provides the determination of the exact energy of pure-spin states on the basis of the obtained single-determinant results within each of the computational schemes at the level of theory that is equivalent to the explicit CI. According to the approach, the energy of pure-spin singlet state is expressed as

$$E^{PS}(0) = E^{U}(0) + S_{max}J. \qquad (4)$$

Where, $E^{U}(0)$ is the energy of the singlet state of the USB solution while $S_{max}$ is the highest spin of the studied odd electron system and $J$ presents the exchange integral

$$J = \frac{E^{U}(0) - E^{U}(S_{max})}{S_{max}^2}. \qquad (5)$$

Here, $E^{U}(S_{max})$ is the energy of the highest-spin-multiplicity state and corresponds to the $S_{max}$-pure-spin state.

Table 2 presents sets of three energies, namely: $E^{R}(0)$, $E^{U}(0)$, and $E^{PS}(0)$, alongside with the exchange integrals $J$ related to ($n_a$,$n_z$) NGrs considered earlier. As seen in the table, comparing with $E^{R}(0)$, the odd electron correlation causes lowering of not only $E^{PS}(0)$ energy, but $E^{U}(0)$ as well, whilst much less pronounced in the latter case, since the pure-spin energy $E^{PS}(0)$ occurs to be the lowest. As seen from the table, the percentage quantities $\delta E^{RPS} = \Delta E^{RPS}/E^{R}(0)$ and $\delta E^{UPS} = \Delta E^{UPS}/E^{U}(0)$, where $\Delta E^{RPS} = E^{R}(0) - E^{PS}(0)$ and $\Delta E^{UPS} = E^{U}(0) - E^{PS}(0)$ present the corresponding energy misalignment, deviate differently: if $\delta E^{RPS}$ changes from ~20 to 25%, $\delta E^{UPS}$ varies much less within ~2-5%. These values clearly show the measure of incorrectness that is introduced when the graphene molecule energy is described by either restricted or unrestricted computational schemes.

Table 2. Energies of singlet ground state and exchange integral of rectangular graphene fragments [*], *kcal/mol* [17]

| Fragment $(n_a, n_z)$ | $E^R(0)$ | $E^U(0)$ | $E^{PS}(0)$ | $\Delta E^{RPS}$ | $\delta E^{RPS}$ [**] % | $\Delta E^{UPS}$ | $\delta E^{UPS}$ [**] % | $J$ |
|---|---|---|---|---|---|---|---|---|
| (5, 5) | 1902 | 1495 | 1432 | 470 | 24.70 | 63 | 4.39 | -1.429 |
| (7, 7) | 2599 | 2223 | 2156 | 443 | 17.03 | 67 | 3.09 | -0.888 |
| (9, 9) | 3419 | 2778 | 2710 | 709 | 20.75 | 68 | 2.53 | -0.600 |
| (11, 10) | 4072 | 3312 | 3241 | 831 | 20.42 | 71 | 2.20 | -0.483 |
| (11, 12) | 4577 | 3676 | 3606 | 971 | 21.22 | 70 | 1.95 | -0.406 |
| (15, 12) | 5451 | 4413 | 4339 | 1112 | 20.40 | 74 | 1.70 | -0.324 |

[*] AM1 version of UHF codes of CLUSTER-Z1. Presented energy values are rounded off to an integer.
[**] The percentage values are related to $\delta E^{RPS} = \Delta E^{RPS}/E^R(0)$ and $\delta E^{UPS} = \Delta E^{UPS}/E^U(0)$, respectively.

## *Answer 2. Broken symmetry approach provides exact determination of the magnetic constant*

Obviously, the odd electrons correlation is a necessary reason for the graphene magnetization. However, this, as such, is not enough since there are additional requirements concerning the magnetic constant value that is equal to the exchange integral $J$ [21] (see Ex (5)). Graphene molecules are among singlet bodies, whose magnetic phenomenon may occur as a consequence of mixing the ground singlet state with those of high-spin multiplicity [22] following, say, to the van Fleck mixing promoted by applied magnetic field [23]. Since the effect appears in the first-order perturbation theory, it depends on the $J$ value that determines the energy differences in denominators. Consequently, $J$ should be small by the absolute value to provide noticeable magnetization. Estimated for molecular magnets [24], the phenomenon can be fixed at $|J|$ of $10^{-2}$-$10^{-3}$ kcal/mol or less.

A joint unit cell of graphene crystal involves two atoms that form one C-C bond of the benzenoid unit. Estimation of $J$ value for ethylene and benzene molecule with stretched C-C bonds up to 1.42Å in length gives -13 kcal/mol and –16 kcal/mol, respectively. In spite of ethylene and benzene molecules do not reproduce the unit cell of graphene crystal exactly; a similar $J$ value of the cell constant is undoubted. Owing to this, magnetization of the graphene crystal cannot be observed so that the crystal should demonstrate the diamagnetic behaviour only.

The latter is supported both theoretically [25] and empirically (see [26] and references therein). To provide a remarkable magnetization means to drastically decrease the magnetic constant $|J|$, which, in its turn, determines a severe strengthening of the odd electron correlation. Since it is impossible to a regular crystal, let us look what can be expected at the molecular level.

Analyzing data published earlier [27, 28] and addressing the discussion presented in the previous section, one may suggest the NGr molecule size as a regulating factor of the electron correlation. As shown in Table 2, the magnetic constant $|J|$ decreases when the molecule becomes larger. When speaking about mixing the ground singlet state with those of high-spin ones, obviously, singlet-triplet mixing is the most influent. As follows from Table 2, the energy gap to the nearest triplet state, equal $2|J|$, for the studied molecules constitutes 2.8-0.6 kcal/mol. The value is still large to provide a recordable magnetization of these molecular magnets, but the trend is quite optimistic.

In view of this idea, let us estimate how large should be graphene molecule to provide a noticeable magnetization. As mentioned earlier, molecular magnetism can be fixed at $|J| \sim 10^{-2}$ -$10^{-3}$ kcal/mol or less. Basing on the data presented in Table 2 and supposing the quantity to be inversely proportional to the number of odd electrons, we get $N \sim 10^5$. For rectangular NGrs with $N$ odd electrons, the number of carbon atoms constitutes $\mathrm{N} = N - 2(n_a + n_z + 1)$ that, according to [19], is determined as

$$\mathrm{N} = 2(n_\alpha n_z + n_\alpha + n_z). \qquad (6)$$

To fit the needed $\mathrm{N}$ value, the indices $n_\alpha$ and $n_z$ should be of hundreds, which leads to linear sizes of the NGrs from a few units to tens *nm*. The estimation is rather approximate, but it, nevertheless, correlates well with experimental observations of the magnetization of activated carbon fibers consisting of nanographite domains of ~2 nm in size [29, 30]. Recently, has been reported a direct observation of size-dependent large-magnitude room-temperature ferromagnetism of graphene interpore regions [31, 32]. The maximum effect was observed at the region width of 20 *nm* after which the signal gradually decreased when the width increased. The behaviour is similar to that obtained for fullerene oligomers [33] that led to the suggestion of a scaly mechanism of nanostructured solid state magnetism of the polymerized fullerene $C_{60}$ and was confirmed experimentally.

The obtained results highlight another noteworthy aspect of the graphene magnetism attributing the phenomenon to size dependent ones. The latter means that the graphene magnetization is observed for nanosize samples only, moreover, for samples whose linear dimensions fit a definite interval, while the phenomenon does not take place at either smaller or bigger samples outside the critical region.

An individual benzenoid unit (including benzene molecule) is non-magnetic (only slightly diamagnetic [34]). When the units are joined to form a graphene-like benzenoid cluster, effectively unpaired electrons appear due to weakening the interaction between odd electrons followed by their correlation. The correlation accelerates when the cluster size increases, which is followed with the magnetic constant $|J|$ decreasing until the latter achieves a critical level that provides a noticeable mixing of the singlet ground state with high-spin states for the cluster magnetization to be fixed. Until the enlargement of the cluster size does not violate a molecular behavior of odd electrons, the sample magnetization will grow. However, as soon as the electron behavior becomes spatially quantized, the molecular character of the magnetization will be broken and will be substituted by that one determined by the electron properties of a crystal unit cell [22]. The critical cluster size is determined by the electron mean free path $l_{el}$. Evidently, when the cluster size exceeds $l_{el}$ the spatial quantization quenches the cluster magnetization. An accurate determination of $l_{el}$ for odd electrons in graphene is not known, but the analysis of a standard data base for electron mean free paths in solids [35] shows the quantity should be ~ 10 *nm*, which is supported by experimental data of 3-7 *nm* electron free path in thin films of Cu-phthalocyanine [36].

Another scenario of getting magnetic graphene is connected with introducing impurity and structural defects in the graphene body. The best illustration of such scenario reality can be found in a recent publication of the Geim team [26] where a paramagnetic behaviour of graphene laminates consisting of 10-50 nm sheets has been recorded after either their fluorination or bombarding by electrons. The treatment provides 'spin-half paramagnetism in graphene induced by point defects'. In both cases, the magnetization is weak and is characterized by one moment per approximately 1,000 carbon atoms, which is explained by the authors by clustering of adatoms and, for the case of vacancies, by the loss of graphene's structural stability. However, the ratio 'one spin per 1,000 carbon atoms' indicates, that, actually, the after-treatment magnetic crystal structure differs from the pristine one and the difference concerns the unit cell that becomes ~33/2 times larger than the previous one. Besides, the unit cell contains one additional spin thus lifting the spin multiplicity to doublet. The latter explains the paramagnetic behaviour of the sample while the size of the cell provides small value of the magnetic constant $|J|$ due to large (~40 *nm*) cell dimension. Therefore, introduced adatoms and point defects cause a magnetic nanostructuring of the pristine crystal that favors the realization of size-dependent magnetism.

Explaining magnetic behavior of graphene molecule, we attribute the phenomenon to the correlation of the molecule odd electrons. As was said in Introduction, criterion 2 highlights the fact that the electron correlation is accompanied with the appearance of effectively unpaired electrons that provide the molecule radicalization [12, 13, 15]. A natural question arises which characteristic of graphene control its electrons correlation? Looking for answering the question we have come to *answer 3*.

## *Answer 3. Odd electrons correlation is controlled by lengths of C-C bonds*

Firstly shown by Takatsuka, Fueno, and Yamaguchi [12], the correlation of weakly interacting electrons is manifested through a density matrix, named as the distribution of 'odd' electrons,

$$D(r|r') = 2\rho(r|r') - \int \rho(r|r'')\rho(r''|r')dr''. \tag{7}$$

The function $D(r|r')$ was proven to be a suitable tool to describe the spatial separation of electrons with opposite spins and its trace

$$N_D = trD(r|r') \tag{8}$$

was interpreted as the total number of these electrons [12, 37]. The authors suggested $N_D$ to manifest the radical character of the species under investigation. Over twenty years later, Staroverov and Davidson changed the term by the 'distribution of *effectively unpaired electrons*' [13, 38] emphasizing that not all odd electrons may be taken off the covalent bonding. Even Takatsuka et al. mentioned [12] that the function $D(r|r')$ can be subjected to the population analysis within the framework of the Mulliken partitioning scheme. In the case of a single Slater determinant, Eq. 8 takes the form [13]

$$N_D = trDS, \tag{9}$$

where

$$DS = 2PS - (PS)^2. \tag{10}$$

Here, $D$ is the spin density matrix $D = P^\alpha - P^\beta$ while $P = P^\alpha + P^\beta$ is a standard density matrix in the atomic orbital basis, and $S$ is the orbital overlap matrix ($\alpha$ and $\beta$ mark different spins). The population of effectively unpaired electrons on atom $A$ is obtained by partitioning the diagonal of the matrix $DS$ as

$$D_A = \sum_{\mu \in A} (DS)_{\mu\mu}, \tag{11}$$

so that

$$N_D = \sum_A D_A. \tag{12}$$

Staroverov and Davidson showed [13] that the atomic population $D_A$ is close to the Mayer free valence index [39] $F_A$ in general case, while in the singlet state $D_A$ and $F_A$ are identical. Thus, a plot of $D_A$ over atoms gives a visual picture of the actual radical electrons distribution [13], which, in its turn, exhibits atoms with enhanced chemical reactivity.

The effectively unpaired electron population is definitely connected with the spin contamination of the UBS solution state. In the case of UBS HF scheme there is a straight relation between $N_D$ and squared spin $\langle S^2 \rangle$ [13]

$$N_D = 2\left( \langle S^2 \rangle - \frac{(N^\alpha - N^\beta)^2}{4} \right), \tag{13}$$

where

$$\langle S^2 \rangle = \left( \frac{(N^\alpha - N^\beta)^2}{4} \right) + \frac{N^\alpha + N^\beta}{2} - \sum_i^{N^\alpha} \sum_j^{N^\beta} |\langle \phi_i | \phi_j \rangle|^2. \tag{14}$$

Here, $\phi_i$ and $\phi_j$ are atomic orbitals; $N^\alpha$ and $N^\beta$ are the numbers of electrons with spin α and β, respectively.

If UBS HF computations are realized in the *NDDO* approximation (the basis for AM1/PM3 semiempirical techniques) [40], a zero overlap of orbitals leads to $S = I$ in Eq. 10, where $I$ is the identity matrix. The spin density matrix $D$ assumes the form

$$D = (P^\alpha - P^\beta)^2. \tag{15}$$

The elements of the density matrices $P_{ij}^{\alpha(\beta)}$ can be written in terms of eigenvectors of the UHF solution $C_{ik}$

$$P_{ij}^{\alpha(\beta)} = \sum_{k}^{N^{\alpha(\beta)}} C_{ik}^{\alpha(\beta)} C_{jk}^{\alpha(\beta)} . \qquad (16)$$

Expression for $\langle \hat{S}^2 \rangle$ has the form [41]

$$\langle \hat{S}^2 \rangle = \left( \frac{(N^\alpha - N^\beta)^2}{4} \right) + \frac{N^\alpha + N^\beta}{2} - \sum_{i,j=1}^{NORBS} P_{ij}^\alpha P_{ij}^\beta . \qquad (17)$$

Within the framework of the *NDDO* approach, the HF-based total $N_D$ and atomic $N_{DA}$ populations of effectively unpaired electrons take the form [42]

$$N_D = \sum_A N_{DA} = \sum_{i,j=1}^{NORBS} D_{ij} \qquad (18)$$

and

$$N_{DA} = \sum_{i \in A} \sum_{B=1}^{NAT} \sum_{j \in B} D_{ij} . \qquad (19)$$

Here, $D_{ij}$ are elements of spin density matrix $D$ that presents a measure of the electron correlation [12, 13, 43], *NORBS* and *NAT* mark the number of orbitals and atoms, respectively.

Explicit expressions (18) and (19) are the consequence of the wave-function-based character of the UBS HF. Since the corresponding coordinate wave functions are subordinated to definite permutation symmetry, each value of spin $S$ corresponds to a definite expectation value of energy [11]. Oppositely, the electron density ρ is invariant to the permutation symmetry. The latter causes a serious spin problem for the UBS DFT [10, 11]. Additionally, the spin density $D(r|r')$ of the UBS DFT depends on spin-dependent exchange and correlation functionals and can be expressed analytically in the former case only [11]. Since the exchange-correlation composition deviates from one method to the other, the spin density is not fixed and deviates alongside with the composition. Serious UBS DFT problems are known as well in the relevance to $\langle \hat{S}^2 \rangle$ calculations [44, 45].

These obvious shortcomings make the UDFT approach practically inapplicable in the case when the correlation of weakly interacting electrons is significant. Certain optimism is connected with a particular view on the structure of the density matrix of effectively unpaired electrons developed by the Spanish-Argentine group [15, 43, 46] from one hand and new facilities offered by Yamagouchi's approximately spin-projected geometry optimization method intensely developed by a Japanese team [47, 48], from the other. By sure, this will give a possibility to describe the electron correlation at the density theory level more thoroughly.

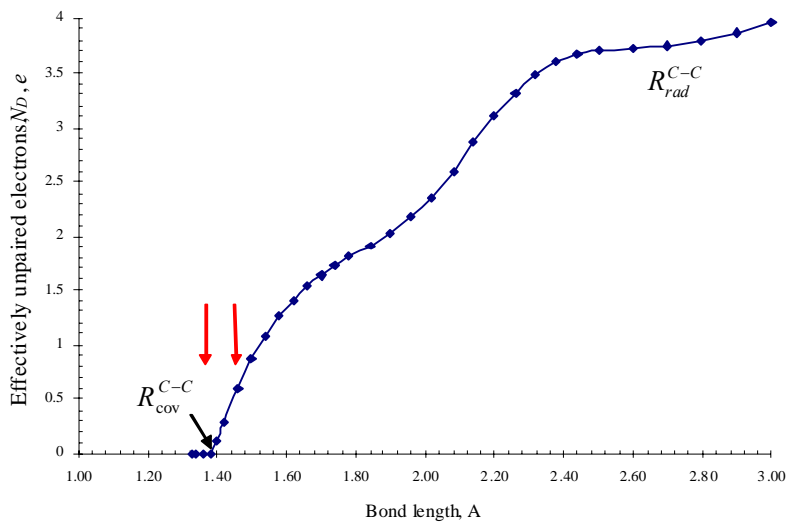

**Fig. 1**. The total number of effectively unpaired electrons $N_D$ accompanying a stretching of C-C bond in ethylene. $R_{cov}^{C-C}$ marks the extreme distance that corresponds to the completion of the covalent bonding. $R_{rad}^{C-C}$ matches a completion of homolytic bond cleavage. Two vertical arrows mark the interval of the C-C bond lengths characteristic for $sp^2$ nanocarbons.

The odd electrons story is counted from the discovery of the benzene molecule made by Michael Faraday in 1825. However, only a hundred years later Hückel suggested the explanation of a deficiency of hydrogen atoms in the molecule to complete the valence ability of carbon atoms. Extra, or odd, electrons were named as $\pi$ electrons that, in contrast to $\sigma$ electrons, interact much weaker while providing an additional covalent coupling between neighbouring atoms. The two electrons are located in the same space and their spins are subordinated to the Pauli law. Formally, this view on extra $\pi$ electrons, which lays in the foundation of the aromaticity concept, has been expanded over all $sp^2$ nanocarbons and has been shared by a number of material scientists in the field until now. However, the concept does not take into account a crucial role of the distance between two neighbouring odd electrons. As seen in Fig. 1, which presents a plotting of the total number of effectively unpaired electrons $N_D$ as a function of the C-C distance in ethylene molecule, the bond stretching from its equilibrium value of 1.326Å up to $R_{crit} = R_{cov}^{C-C} = 1.395$Å does not cause the appearance of the unpaired electrons so that the relevant π electrons are fully covalently bound. However, above $R_{crit}$ the number $N_D$ gradually increases up to a clearly vivid knee that is characterized by $N_D \cong 2$ at $R$=1.76Å, which evidences a complete radicalization of the previous π electrons. On the way from $R_{crit}$ to $R$=1.76Å the two electrons are not more located in the same space, but electrons with different spins occupy different spaces. Further stretching concerns mainly two σ electrons that, once fully covalently bound until $R$=1.76Å, gradually become unpaired just repeating the fortune of π electrons resulting in $N_D \cong 4$ at 2.5Å.

In spite of clear explanation where unpaired electrons are coming from, the question about their existence still remains due to suspicion of their attribution to an *artifact* caused by the limitations of single-determinant calculations. Looking for confirmation of physical reality of unpaired electrons leads to *answer 4*.

## *Answer 4. Effectively unpaired electrons are a definite physical reality*

In a series of aromatic hydrocarbon molecules, the unified length of C-C bonds in the benzene molecule exactly fits $R_{crit}$, which is why $N_D$=0 as is expected for a truly aromatic molecule. However, even naphthalene molecule is characterized by a set of C-C bonds, short and long representatives of which have lengths that are below and above $R_{crit}$, respectively. This slightly dispersive many-length set is further kept in all aromatic molecules (becoming a two-length one in fullerene $C_{60}$). As the number of benzene units grows, the number of long bonds increases, which is followed by increasing $N_D$ (see Table 3) [16]. As seen in the table, for pentacene, $N_D$ constitutes 5.4 *e* so that the molecule is a 5.4-fold radical. The $N_D$ distribution over the molecule atoms in terms of $N_{DA}$ is shown in Fig. 2a. As seen in the figure, the main chemical reactivity of the molecule is concentrated in its central part.

This finding could have been one of questionable results of the molecular theory only if it were not for a recent experimental viewing of the molecule by using AFM with unprecedented high resolution [49] (see Fig.2b). The molecule image was obtained by using short-range chemical forces of noncontact AFM. The forces profiles are shown in Fig. 2c. As seen in the figure, the least forces and, consequently, the weakest interaction are observed at the molecule ends (the brightest area in Fig. 2b) while the strongest interaction (the darkest area in Fig. 2b) is characteristic for the molecule central area. Since the interaction of the CO apex of the AFM tip is obviously proportional to the electron density on the atoms above which the tip is located, thus recorded AFM molecule image should be inverted by color with respect to the $N_{DA}$ image map in Fig. 2a. This has actually been observed exhibiting the first evidence of the distribution of effectively unpaired electrons in $sp^2$ molecules. The next example concerns a similar imaging of olympicene molecule that has been synthesized on the eve of the London Olympic Games 2012 [50]. Figure 3 presents the image map of the $N_{DA}$ distribution over the molecule alongside with its AFM image obtained as previously. The color inversion of the two images is clearly seen.

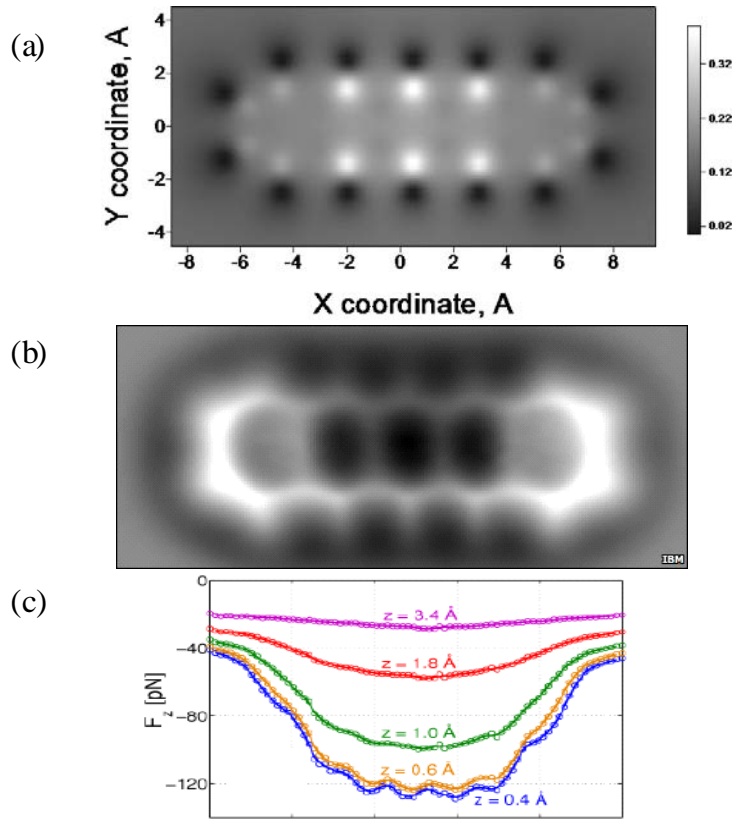

**Fig. 2**. Effectively unpaired electrons of a pentacene molecule. a. Calculated $N_{DA}$ image, UBS HF singlet state. b. AFM imaging of pentacene on Cu(111) using a CO-at-Au tip [49]. c. Extracted vertical force along the molecule long axis [49].

**Table 3.** Effectively unpaired electrons in aromatic molecules, UBS HF singlet state [16]

| Molecules | C-C bond length, E Number of bonds | | | | $N_D$ |
|---|---|---|---|---|---|
| Benzene | 1.395 6 | | | | 0.05 |
| Naththalene | 1.385 4 | 1.411 2 | 1.420 4 | 1.430 1 | 1.483 |
| Anthracene | 1.387 4 | 1.410 6 | 1.421 4 | 1.435 2 | 3.003 |
| Tetracene | 1.388 4 | 1.410 8 | 1.421 6 | 1.436 3 | 4.320 |
| Pentacene | 1.388 4 | 1.411 10 | 1.420 8 | 1.436 4 | 5.540 |

(a)
(b)

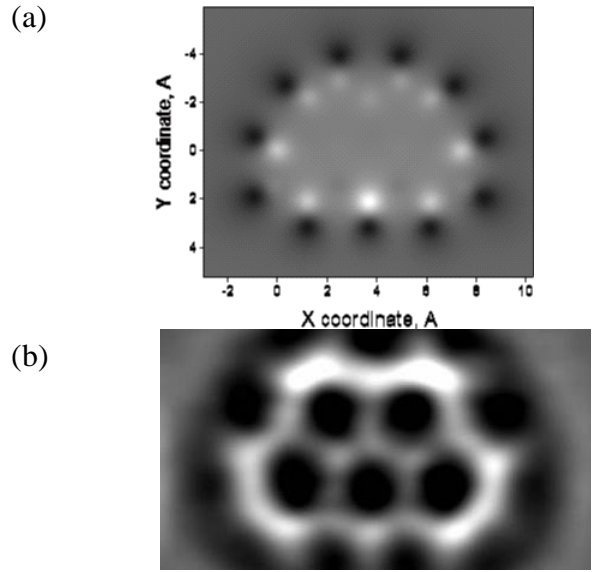

**Figure 3**. Effectively unpaired electrons of an olympicene molecule. Calculated $N_{DA}$ image, UBS HF singlet state (a) and AFM imaging of olympicene on Cu(111) using a CO-at-Au tip [50].

Two vertical arrows in Fig.1 mark the C-C bond length interval that is characteristic for graphene molecules equilibrated in the framework of the UBS HF approach. (It should be mentioned that application of the restricted version of the same program results in practically non-dispersive value of C-C bond length of 1.42Å.) As seen, C-C bond lengths exceed $R_{crit}$ which leads to a considerable amount of effectively unpaired electrons, total numbers of which are listed in Table 1 for different graphene fragments. Figure 4a exhibits a typical image map of the $N_{DA}$ distribution over one of them. The fragment edges are not terminated and the $N_{DA}$ image map has a characteristic view with a distinct framing of the sample by edge atoms since the main part of the unpaired electrons are concentrated in this area. The $N_{DA}$ image map intensity in the basal plane is of ~0.3 $e$ in average.

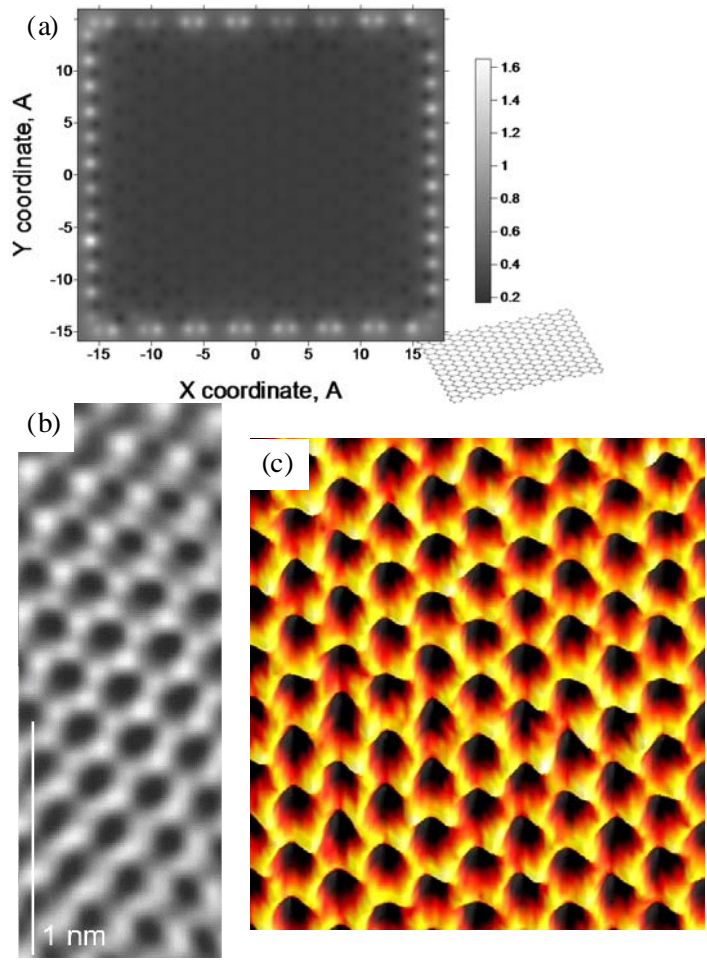

**Fig. 4**. Effectively unpaired electrons of graphebe. a. Calculated $N_{DA}$ image of (15, 12) NGr molecule, UBS HF singlet state. b. An atomic resolution image of a portion of a graphene monolayer [53]. c. The image of a single suspended sheet of graphene taken with the TEAM 0.5, at Berkeley Lab's National Center for Electron Microscopy [54]

The peculiarity of the graphene edges has been a topic for intense discussion from the very beginning of the graphene science [51] when they were disclosed by using a tight-binding band calculation within the Hückel approximation [52]. However, they have not been attributed to the effectively unpaired electrons and have been discussed in the context of the graphene spin system peculiarity with respect to expected magnetic behavior of the sample. In this context, it is worthwhile to refer to one more quote from the Hoffmann 'informal reflection' [1]: "There is a special problem that theory has with unterminated structures—ribbons cut off on the sides, polymers lacking ends. If passivation is not chosen as a strategy, then the radical lobes of the unterminated carbon atoms, or undercoordinated transition metals, will generate states that are roughly in the middle energetically, above filled levels, below empty levels in a typical molecule that has a substantial gap between filled and unfilled levels. If such levels—states, the physicists call them—are not identified as "intruder" states, not really real, but arising from the artifact of termination, they may be mistaken for real states in the band gap, important electronically. And if electrons are placed in them, there is no end to the trouble one can get into. These band gap states are, of course, the origin of the reactivity of the terminated but not passivated point, line, or plane. But they have little to do with the fundamental electronic structure of the material". Supporting the said above, depicted in Fig. 4a presents the reactivity image of the graphene molecule. As seen in the figure, not only edge, but basal-plane carbon atoms are chemically active, albeit with different efficacy. Important to note that the reactivity is rather inhomogeneous. The recent atom-resolved graphene images convincingly witness this inhomogeneity as can be seen in Figs. 4b and c. Therefore, effectively unpaired electrons of $sp^2$ molecules are a physical reality and are assuming their leading place in the molecular theory of graphene.

In the singlet state, the $N_{DA}$ values are identical to the atom free valences [13] and thus exhibit the atomic chemical susceptibility (ACS) [55, 56]. The $N_{DA}$ distribution over atoms plots a 'chemical portrait' of the studied molecule, whose analysis allows for making a definite choice of the target atom with the highest $N_{DA}$ value to be subjected to chemical attack by an external addend. Therefore, we have come to *answer 5* claiming that peculiarities of the graphene chemistry can be exhibited at a quantitative level, much as this has been done for fullerenes [5].

## *Answer 5. Computational strategy of the chemical modification of graphene*

A typical chemical portrait of graphene fragment highlights edge atoms as those with the highest chemical activities, besides rather irregular, while exhibiting additionally the basal atoms ACS comparable with that one of fullerene $C_{60}$ [57, 58].

This circumstance is the main consequence of the odd electron correlation in graphene in regards its chemical modification. Ignoring the correlation has resulted in a common conclusion about chemical inertness of the graphene atoms with the only exclusion concerning edge atoms. Owing to this, a computationist does not know the place of both the first and consequent chemical attacks to be possible and has to perform a large number of calculations sorting them out over the atoms by using the lowest-total-energy (LTE) criterion (see, for example, [59]). In contrast, basing of the $N_{DA}$ value as a quantitative pointer of the target atom at any step of the chemical attack, one can suggest the algorithmic 'computational syntheses' of the molecule polyderivatives [60]. In what follows the algorithm-in-action will be illustrated by the examples of the hydrogenation and oxidation of the (5, 5) NGr molecule.

**(5, 5) NGr molecule hydrogenation**

The equilibrium structure of the (5,5) NGr molecule alongside with its $N_{DA}$ image map is shown in Fig.5. Panel *b* exhibits the $N_{DA}$ distribution attributed to the atoms positions thus presenting the 'chemical portrait' of the molecule. Different $N_{DA}$ values are plotted in different colouring according to the attached scale. The absolute $N_{DA}$ values are shown in panel *c* according to the atom numbering in the output file. As seen in the figure, 22 edge atoms involving 2x5 *zg* and 2x6 *ach* ones have the highest $N_{DA}$ thus marking the perimeter as the most active chemical space of the fragment. The hydrogenation of the fragment will start on atom 14 (star-marked in Fig.5c) according to the highest $N_{DA}$ in the output file. The next step of the reaction involves the atom from the edge set as well, and this is continuing until all the edge atoms are saturated by a pair of hydrogen atoms each since all 44 steps are accompanied with the high-rank $N_{DA}$ list where edge atoms take the first place. Thus obtained hydrogen-framed graphene molecule is shown in Fig. 6 alongside with the corresponding $N_{DA}$ image map. Two equilibrium structures are presented. The structure in panel *a* corresponds to the optimization of the molecule structure without any restriction. In the second case, positions of edge carbon atoms and framing hydrogen atoms were fixed and the optimization procedure results in the structure shown in panel *c*. In what follows, we shall refer to the two structures as free standing and fixed membranes, respectively. Blue atoms in Fig. 6c alongside with framing hydrogens are excluded from the forthcoming optimization under all steps of the further hydrogenation.

      Chemical portraits of the structures shown in Fig. 6b and Fig. 6d are quite similar and reveal the transformation of brightly shining edge atoms in Fig. 5b into dark spots. The addition of two hydrogen atoms to each of the edge ones saturates

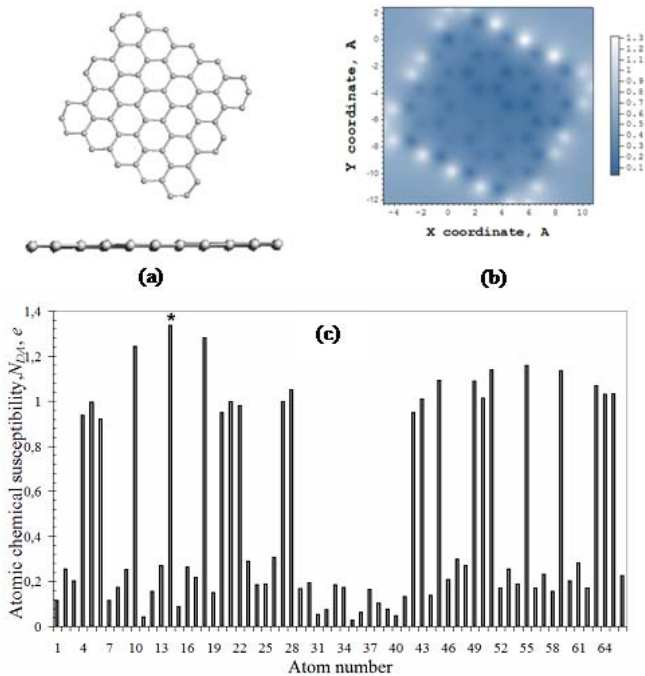

**Fig. 5**. Top and side views of the equilibrium structure of (5,5) NGr molecule (a); $N_{DA}$ image map (b) and $N_{DA}$ distribution over atoms according to atom numbers in the output file (c) [61]

the valence of the latter completely, which results in zeroing $N_{DA}$ values, as is clearly seen in Fig. 6e. The chemical activity is shifted to the neighbouring inner atoms and retains higher in the vicinity of *zg* edges, however, differently in the two cases. The difference is caused by the redistribution of C-C bond lengths of free standing membrane when it is fixed over perimeter, thus providing different starting conditions for the hydrogenation of the two membranes.

Besides the two types of initial membranes, the hydrogenation will obviously depend on other factors, such as 1) the hydrogen species in use and 2) the accessibility of the membranes sides to the hydrogen. Even these circumstances evidence the hydrogenation of graphene to be a complicated chemical event that strongly depends on the initial conditions, once divided into 8 adsorption modes in regards atomic or molecular adsorption; one- or two-side accessibility of membranes; and free or fixed state of the membranes perimeter. Only two of the latter correspond to the experimental observation of hydrogenated specimens discussed in [62], namely: two-side and one-side atomic hydrogen adsorption on the fixed membrane. Stepwise hydrogenation of the (5, 5) NGr molecule was considered in details in [61]. Here, we restrict ourselves with a brief description of the main results.

*Two-side atomic adsorption of hydrogen on fixed membrane.* The hydrogenation concerns the basal plane of the fixed hydrogen-framed membrane shown in Fig. 6c that is accessible to hydrogen atoms from both sides. As seen in Fig. 6e, the first hydrogenation step should occur on basal atom 13 marked by a star. Since the membrane is accessible to hydrogen from both sides, one has to check which deposition of the hydrogen atom, namely, above the carbon plane ('up') or below it ('down') satisfies the LTE criterion.

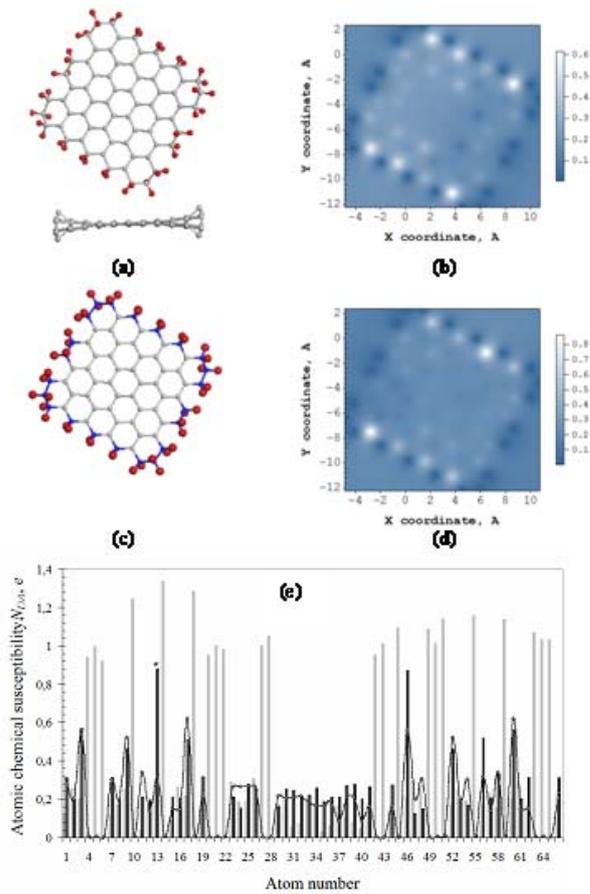

**Fig. 6**. Equilibrium structures of free standing (top and side views) (a) and fixed (c) (5,5) NGr membrane; $N_{DA}$ image maps (b, d) and $N_{DA}$ distribution over atoms according to atom numbers in the output file (e) [61]. Light gray histogram plots ACS data for the pristine (5,5) NGr molecule. Curve and black histogram are related to membranes in panels *a* and *c*, respectively.

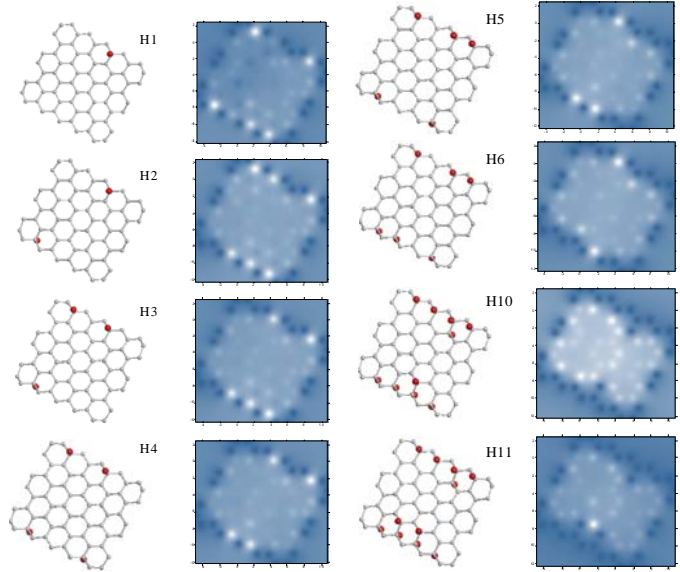

**Fig. 7.** Equilibrium structures (left) and $N_{DA}$ image maps (right) of graphene hydrides 1 related to initial stage of the basal-plane hydrogenation. HKs denote hydrides with K hydrogen atoms deposited on the membrane basal plane [61]. Framing hydrogen atoms are not shown to simplify the structure image presentation.

After deposition of hydrogen atom on basal atom 13, the $N_{DA}$ map has revealed carbon atom 46 for the next deposition (see H1 $N_{DA}$ map in Fig. 7). The LTE criterion favours the down position for the second hydrogen on this atom so that we obtain structure H2 shown in Fig. 7. The second atom deposition highlights next targeting carbon atom 3 (see $N_{DA}$ map of H2 hydride), the third adsorbed hydrogen atom activates target atom 60, the fourth does the same for atom 17, and so forth. Checking up and down depositions in view of the LTE criterion, a choice of the best configuration can be performed and the corresponding equilibrium structures for a selected set of hydrides from H1 to H11 are shown in Fig. 7. The structure obtained at the end of the 44[th] step is shown in Fig. 8a. It is perfectly regular, including framing hydrogen atoms thus presenting a computationally synthesized fully saturated chairlike (5, 5) NGr polyhydride that is in full accordance with the experimental observation of the graphane crystalline structure [62].

*One-side atomic adsorption of hydrogen on fixed membrane.* Coming back to the first step of the hydrogenation, which was considered in the previous Section, let us proceed further with the second and all the next steps of the up deposition only. As previously, the choice of the target atom at each step is governed by high-rank $N_{DA}$ values. Figure 8b presents the saturated graphene polyhydride related to the final 44$^{th}$ step. A peculiar canopy shape of the carbon skeleton of the hydride is solely by the formation of the table-like cyclohexanoid units. However, the unit packing is only quasi-regular that may explain the amorphous character of the hydrides formed at the outer surface of graphene ripples observed experimentally [62]. The reasons of the hydrogen molecule desorption at the step are discussed elsewher [61]. A complete set of the one-side obtained products form hydrides family 2.

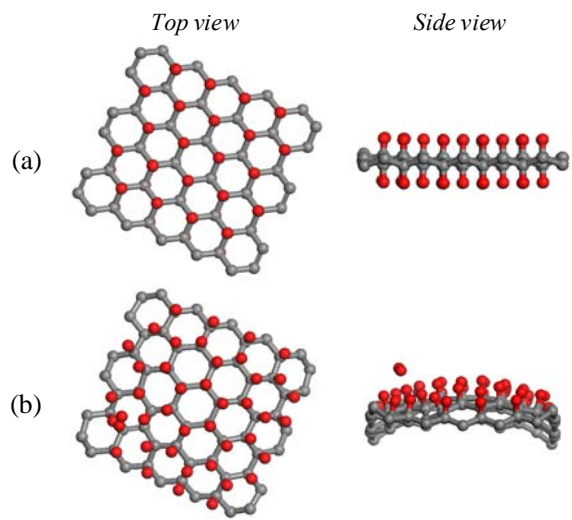

**Fig. 8.** Top and side views of the equilibrium structures of the saturated graphene hydrides formed at the atomic adsorption of hydrogen on the fixed (5,5) NGr membrane, accessible to the adsorbate from both (a) and one (b) sides [61].

As for the hydrogen coverage, Fig. 9 presents the distribution of C-H bond lengths of saturated graphene polyhydrides of families 1 and 2. In both cases, the distribution consists of two parts, the first of which covers 44 C-H bonds formed at the molecule skeleton edges. Obviously, this part is identical for both hydrides since the bonds are related to framing atoms. The second part covers C-H bonds formed by hydrogen atoms attached to the basal plane. As seen in the figure, in the case of hydride 1, C-H bonds are practically identical with the average length of 1.126Å and only slightly deviate from those related to framing atoms. This is just a reflection of the regular graphane-like structure of the hydride shown in Fig. 8a similarly to highly symmetric fullerene hydride $C_{60}H_{60}$ [63]. In contrast, C-H bonds on a canopy-like carbon skeleton of hydride 2 are much longer than those in the framing zone, significantly oscillate around the average value of 1.180Å. In spite of the values greatly exceed a 'standard' C-H bond length of 1.11Å, typical for benzene, those are still among chemical C-H bonds, whilst stretched, since the C-H bond rupture occurs at the C-H distance of 1.72Å

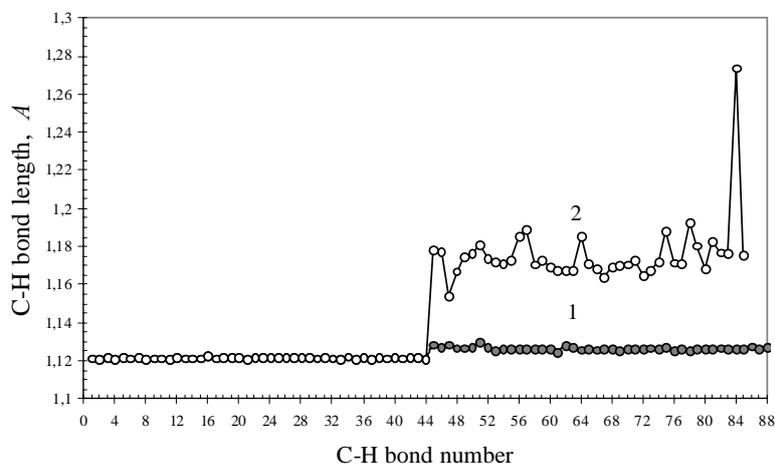

**Fig. 9**. C-H bond length distribution for saturated graphene polyhydrides of families 1 (1) and 2 (2) [61].

[64]. A remarkable stretching of the bonds points to a considerable weakening of the C-H interaction for hydrides 2, which is supported by the energetic characteristics of the hydrides, as well [61]. The total energies of both hydrides are negative by sign and gradually increase by the absolute value when the number of adsorbed atoms increases. Besides, the absolute value growth related to hydrides 2 is evidently slowing down starting at step 11 in contrast to the continuing growth for hydrides 1. This retardation obviously shows that the addition of hydrogen to the fixed membrane of hydrides 2 at coverage higher than 30% is more difficult than in the case of hydrides 1. The reaction of the chemical attachment of hydrogen atoms for hydrides 1 is thermodynamically profitable through over the covering that reaches 100% limit. In contrast, the large coverage for hydrides 2 becomes less and less profitable so that at final steps adsorption and desorption become competitive.

**(5, 5) NGr molecule oxidation**

Stepwise oxidation of the (5, 5) NGr molecule has been considered similarly to the hydrogenation described in the previous section. On the background of a tight similarity in both processes, in general, important difference of the events concerns the fact that instead of atomic hydrogens which were attacking agents in the first case, a set of oxidants consisting of oxygen atoms O, hydroxyls OH, and carboxyls COOH had to be considered in the latter case. A detailed description of the molecule oxidation is given in [65, 66]. Skipping extended explanations of details which were given above for hydrogenation, a brief presentation of results of the performed computational experiment, attributed to main hot points of the graphene oxide (GO )chemistry that have been still under question, can be presented in the following way.

*Morphology*. Empirical experiments reveal a remarkable disordering of the initial graphene structure even by partial oxidation so that chemically produced GOs are highly amorphous (see [67-70] and references therein).

None of the regular structured GO has been obtained in the study. Therefore, the performed computational experiment fully supports this finding and forces to think about high flexibility of the graphene structure and its highly sensitive response to any external action.

*Graphene oxidation as a process in general*. Experimentally was shown that the oxidation of the graphene proceeds in a rather random manner [67]. The saturated at% ratio of oxygen to carbon is ~0.20-0.45 [70-73]. When graphene oxide is heated to $1100^0$ C, there is still about 5-10 at % oxygen left [72-74].

All the features are supported by the calculations. As shown, the oxidation can be considered as a stepwise addition of oxidants to the pristine graphite body while the addition sequences for each monolayer are subordinated to a particular algorithm governed by the list of high-rank atomic chemical susceptibilities $N_{DA}$. In numerous cases presented in [65, 66], was shown that the algorithm action does cause seemingly random distribution of oxidants over the pristine body in due course of the oxidation process.

The algorithmic approach does not impose any restriction on the limit at% ratio of any addend to carbon, in general. This was supported by the results of the 'computational synthesis' of polyderivatives of fullerene $C_{60}$ [5] as well as polyhydrides and polyfluorides of the (5, 5) NGr molecule [61]. However, the performed computations, in full consent with previously considered hydrogenation of the molecule, have revealed that the initial radicalization of the molecule, which is provided by $N_D$ effectively unpaired electrons, has been gradually suppressed as the chemical reaction proceeds. The molecule chemical reactivity is gradually worked out approaching zero due to which the reactions stop. This explains why the hydrogenation and fluorination of fullerene $C_{60}$ is terminated at producing $C_{60}F_{48}$ and $C_{60}H_{36}$ polyderivatives, respectively, [60, 75] and why at% ratio of hydrogen to carbon in experiment of Elias et al. [62] is less than 133.3% when going from hydrides formed from two-side H-accessible perimeter fixed graphene membranes to one-side H-accessible graphene ripples [61]. The same regularities govern the graphene molecule oxidation, which, as shown, terminates the oxidation at achieving 112at% and 65at% of oxygen when the oxidation is provided by addition of either hydroxyls or oxygen atoms only. The latter situation is shown to be much more preferential. The saturation number involves filling both edge and basal atoms. Since the model molecule is rather small, the contribution of edge atoms is significant. If exclude this contribution, the number of 48at% is characteristic for (5, 5) NGr GO that is quite reasonable and points to a predominant $C_2O$ stoichiometry on the basal plane. Since experimental samples are much bigger in size, the mentioned earlier data of ~20-45at% are mainly related to the basal positions since the contribution of edge atoms is ≤1% at linear dimension of ≤0.1μ.

In contrast, the availability of remaining oxygen in reduced GOs (rGOs) subjected to heating up to $1100^0$ C, is connected with edge atoms of the latter. As shown, these atoms, which include not only perimeter atoms of the rGOs molecules but the atoms framing every defect zone, form a local area with very high chemical reactivity. Oxidants are strongly coupled with the atoms and can leave the molecule only alongside with carbon partners. The number of such atoms depends on linear size of both pristine GO molecules and their inner defects and cannot evidently exceed a few percents, which perfectly correlates with the observed amount of the remaining oxygen.

*Chemical composition of graphene oxide.* Basing on empirical data, the most common opinion attributes COOH, OH, and C=O groups to the edge of the GO sheet, while the basal plane is considered to be mostly covered with epoxy C-O-C and OH groups [67, 70, 76, 77].

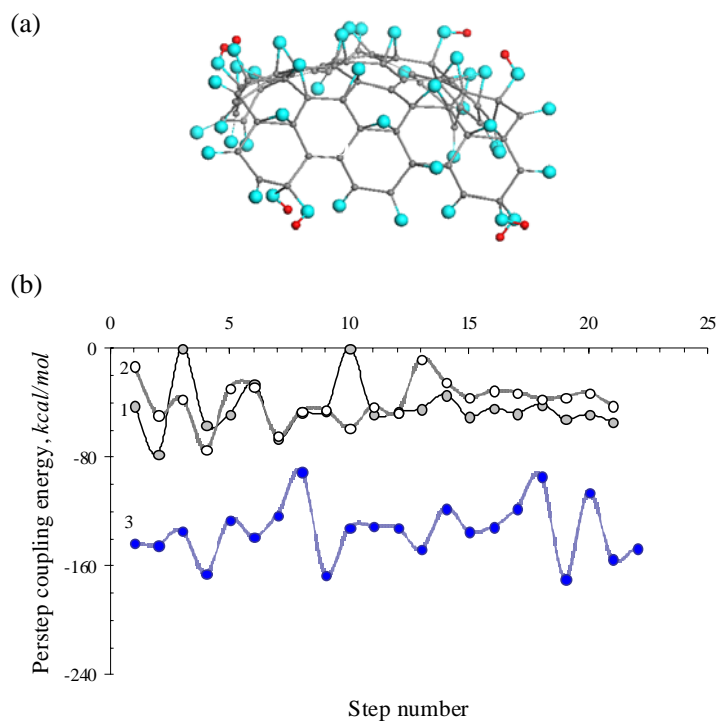

**Fig.10**. Based on the (5,5) NGr molecule, structural model of a top-down exfoliated GO (a) and per step coupling energy (b) versus step number for this GO family under subsequent O- and OH-additions to carbon atoms at either the molecule basal plane (curves 1 and 2) or edges (curve 3).

The performed computations have allowed for forming up a hierarchy of the main three oxidants (O, OH, COOH) in regards their participation in the graphene oxidation that has shown an extremely low probability of such activity for carboxyls. Basing on the results obtained, it is possible to suggest a reasonable, self-consistent model of a convenient GO presented in Fig.10a. Sure, the model cannot be simply scaled for adapting to larger samples. Obviously, due to extreme sensitivity of the graphene molecule structure and electronic system to even small perturbations caused by external factors, the fractional contribution of O, OH, and C-O-C groups may change in dependence of changing the molecule size, shape as well as of the presence of such impurities as metal atoms [78] and so forth. These facts may explain 'fluidness' of the term "graphene oxide" pointed by Ruoff et al [67]. However, it is possible to convincingly state that the chemical composition of any GO has been governed by the presence of two zones drastically differing by the coupling of the relevant oxidants with the graphene molecule body so that carbonyl/hydroxyl and epoxy/hydroxil combinations will be typical for edge and basal areas of all GOs of different size and shape (see Fig. 10b).

Besides the chemical composition of chemically produced GOs, the performed calculations are able to suggest the chemical composition of rGOs as well. Discussion based on a two-zone-chemical-reactivity peculiarity of graphene molecules, clearly pointed to a reliable rGO model shown in Fig. 11.

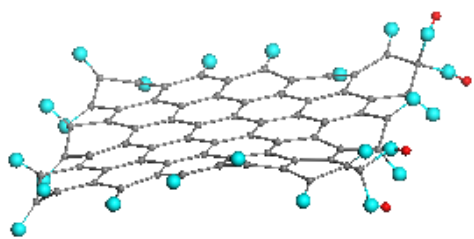

**Fig. 11**. Based on the (5,5) NGr molecule, structural model of a reduced top-down exfoliated rGO [65, 66].

Concluding discussion of hydrogenation and oxidation of graphene, some words should be said concerning the computational strategy applicable to the molecule chemical modification, in general. Until now, the computational strategy in this field has been aimed at finding support to one of the available models, the majority of which has been suggested just intuitively. This strategy has been a result of certain limitations provided by a standard computational DFT scheme within the framework of solid-state periodic boundary conditions that requires a beforehand given structure of the relevant supercell unit. However, the computational study, based on such concept 'from a given structure to reliable properties' has resulted in wrong conclusions, which, for example, in the case of GO have led to the statement about kinetically constrained metastable nature of GO [79], thus revealing its inability to meet calls of the GO chemistry. In contrast, the molecular theory of graphene does not need any given structure beforehand but creates the structure in due course of calculations following the algorithms that take into account such fragile features of graphenes as their natural radicalization, correlation of their odd electrons, an extremely strong influence of structure on properties, a sharp response of the graphene molecule behavior on small action of external factors.

Molecular theory not only well works with graphene chemical modification but opens large possibility in considering mechanical properties of graphene, in general, and its mechanochemistry, in particular, thus suggesting *answer 6*.

## *Answer 6. Electron correlation of graphene is strongly influenced by mechanical deformation*

Molecular theory considers graphene deformation in terms of mechanochemical reactions which allows disclosing very deeply rooted peculiarities in the molecule mechanical behavior, invisible at first glance. Deformation of graphene is tightly connected with odd electron correlation since it concerns changing interatomic distances. As we saw, the latter is very important regulator of the correlation extent thus increasing it when the distance grows. Obviously, strengthening of electron correlation results in the growths of the number of effectively unpaired electrons $N_D$ as it was shown in Fig. 1.

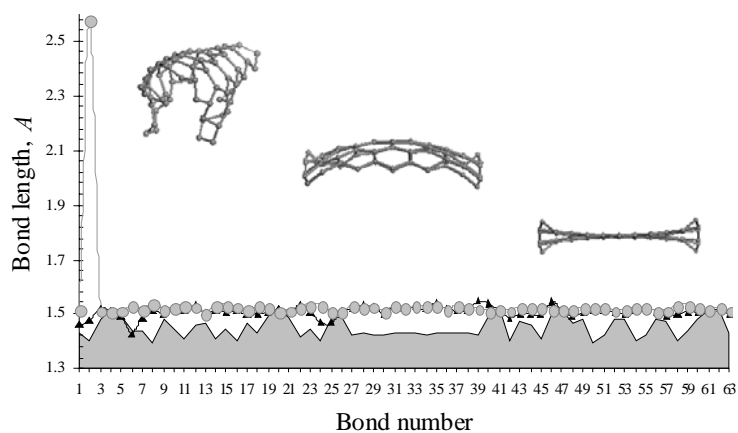

**Fig. 12**. C-C bond length distribution for carbon skeletons of the pristine (5, 5) NGr (gray filled region), canopy-like (dark triangles) and basket-like (gray balls) fixed membranes [61].

A similar stretching of C-C bonds can be highlighted when comparing the carbon skeletons of the pristine (5, 5) NGr molecule and those canopy-like and basket-like ones subjected to one-side hydrogen adsorption on either fixed or free standing membrane, respectively [61]. Figure 12 presents the views of the skeletons alongside with the distribution of their C-C bond lengths. As seen in the figure, C-C bonds of both deformed skeletons are elongated, whilst the summary elongation for the basket-like skeleton is evidently bigger than that one for the canopy-like one. The elongation is restricted by the bond length 1.53Å, which is dictated by $sp^3$ configuration of carbon atoms due to hydrogenation. Naturally, the accumulated deformation may cause some bonds breaking, which occurs for bond 2 of the basket-like skeleton. As a whole, changes in the C-C bond lengths presented in Fig.12 result in decreasing magnetic constant $J$ by the absolute value from -1.43 kcal/mol for the pristine (5, 5) NGr to -0.83 and -0.59 kcal/mol for the canopy-like and basket-like skeletons. Simultaneously, $N_D$ increases from 31$e$ to 46$e$ and 54$e$, respectively. Both findings evidence an undoubted strengthening of the odd electron correlation caused by the chemically-stimulated deformation of the carbon skeleton.

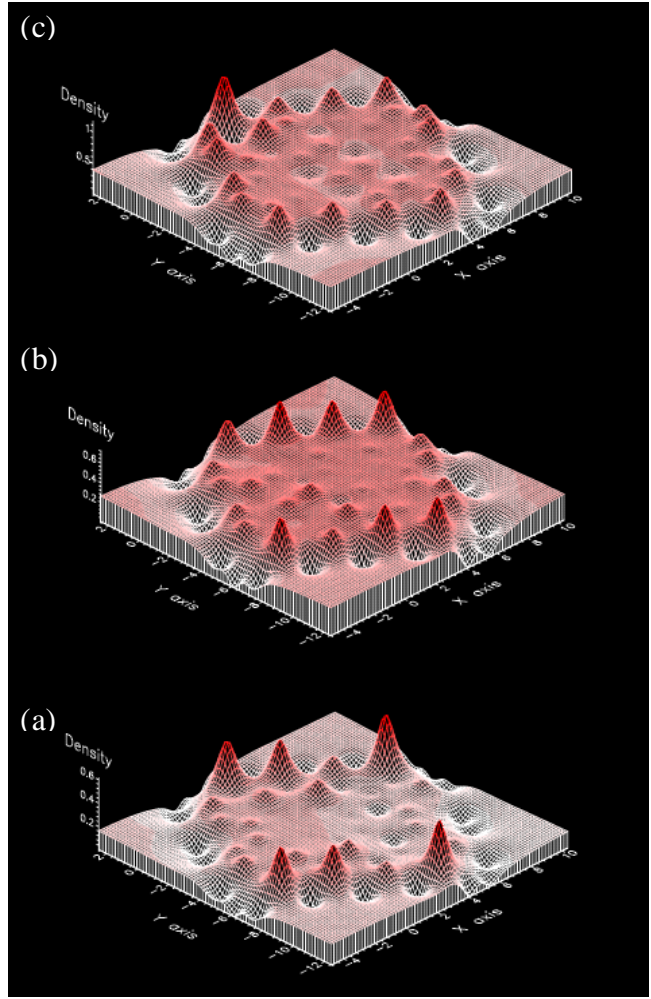

**Fig. 13.** Effectively-unpaired-electron-density images of the carbon skeletons of the pristine (5, 5) NGr (a), canopy-like (b) and basket-like (c) fixed membranes.

Yet another evidence of the deformation effect is presented in Fig.13. The figure shows the redistribution of unpaired electrons density over the skeleton atoms caused by the deformation. As seen in the figure, the skeleton electron-density image greatly changes when the electron correlation becomes stronger (draw attention on a large vertical scale of plottings presented in the top figure). Consequently, if observed by HRTEM, the basket-like skeleton might have look much brighter than the canopy-like one and especially than the least bright pristine molecule. In view of finding, it is naturally to suggest that raised above the substrate and deformed areas of graphene in the form of bubbles, found in a variety of shapes on different substrates [80, 81], reveal peculiar electron-density properties just due to stretching deformation that results in strengthening the odd electron correlation. Small (5, 5) NGr molecule presented in Fig. 13 cannot pretend to simulate the picture observed for micron bubbles, but it exhibits a general trend that might take place in bubbles, as well. In view of the obvious strengthening of the odd electron correlation caused by the deformation, this explanation looks more natural than that proposed from the position of an artificial 'gigantic pseudomagnetic field' [80].

Abovementioned considerable decreasing of magnetic constants $J$ stimulated by deformation allows for suggesting a peculiar magnetic behaviour of the deformed graphene regions, such as, say, bubbles, stimulated by both their size and curvature. The two parameters obviously favour decreasing in the constant values thus promoting the appearance of magnetic response localized in the bubble regions.

Besides the formation of bubbles caused by ultrastrong adhesion of graphene membranes to different substrates [82], the deformation of graphene can be caused by the application of external stress. Quantum molecular theory suggests considering the graphene deformation and rupture in terms of a mechanochemical reaction [83-85]. The quantum chemical realization of the approach is based on the coordinate-of-reaction concept for the purpose of introducing a mechanochemical internal coordinate (MIC) that specifies a deformational mode. The related force of response is calculated as the energy gradient along the MIC, while the atomic configuration is optimized over all of the other coordinates under the MIC constant-pitch elongation. When applied to the description of the deformation of both (5, 5) nanographene [83, 84] and (5, 5) nanographane [85] molecules under uniaxial tension, the calculations highlighted a pronounced changing in the number of effectively unpaired electrons $N_D$ of the sample in due course of its deformation. As shown, changing is different when the deformation occurs either along or normal to the chains of C-C bonds. However, in all cases the changing is quite significant pointing to a considerable strengthening of odd electron correla-tion due to changes in interatomic spacings. A detailed consideration of a possible regulating mission of stress with respect to the enhancement of chemical reactivity of carbon atoms and magnetic behaviour of the loaded sample obviously deserves a further thorough study.

## Discussion and conclusive remarks

Odd electrons of benzenoid units and correlation of these electrons having different spins are the main concepts of the molecular theory of $sp^2$ nanocarbons. In contrast to the theory of aromaticity, the molecular theory accepts that odd electrons with different spins occupy different places in the space so that the configuration interaction (CI) becomes the central point of the theory. Consequently, a multi-determinant presentation of the wave function of the system of weakly interacting odd electrons is absolutely mandatory on the way of the theory implementation at the computational level. However, the efficacy of the available CI computational techniques is quite restricted in regards large polyatomic systems, which does not allow performing extensive computational experiments. On the other hand, the modern computational science of $sp^2$ nanocarbons, in general, and graphene, in particular, is, actually, the field of such experiments due to its steadily grown importance caused by prevailing computations over other empirical technique, which is evidently the case of graphene. Facing the problem, computationists have addressed standard single-determinant ones albeit not often being aware of how correct are the obtained results. The current paper is an attempt to present the molecular theory of graphene in terms of single-determinant computational schemes as well as to analyze the reliability of the obtained results.

    Open-shell presentation of wave functions is the first step towards multi-determinant computational schemes so that naturally one has to address this form of the function presentation. Unrestricted Hartree-Fock (UHF) and density functional techniques (UDFT) are to be the basic grounds for the techniques used. In spite of a partial suiting of both approaches to the CI ones, both UHF and UDFT schemes provide spin-contaminated solutions with the relevant energies that exceed the pure-spin ones. Much higher energies and, thus, much less reliability correspond to standard computational HF and DFT schemes in the restricted close-shell approach. Nevertheless, a predominant majority of DFT computations related to graphene have been performed in this very approximation, which impugns greatly the reliability of the results obtained.

    In the case of unrestricted approach, the situation is better but this does not remove the issue about the result reliability. An example of applying the UHF-based theory to graphene, were obtained answers to most of the questions. These answers led the foundation of the current paper. It should be noted that getting

them has required the performance of system computational experiments in the majority of cases.

Before passing to the answers, one should pay attention to the fact that the inner features of the unrestricted computational schemes, which make them different from the exact CI consideration, open the possibility in issuing three computational criteria that can distinguish electrons systems by the electrons correlation. These criteria are presented by the following quantities: 1) the energy misalignment $\Delta E^{RU} \geq 0$; 2) the total number of effectively unpaired electrons $N_D \neq 0$; and 3) the squared spin misalignment $\Delta S^2 \geq 0$. A detailed description of the values is given in the relevant Section. When all the quantities are zero, the electrons are non-correlated (that is the case of benzene molecule) and the relevant *sp²* systems subordinate to the theory of aromaticity. In the case of graphene, the values are not zero, which manifests a considerable correlation of its odd electrons. Studying the graphene odd electrons system by using the unrestricted broken symmetry approach, one can obtain the following answers concerning the issue mentioned above.

Answer 1 states that application of both UHF and UDFT techniques in the framework of the broken symmetry approach [20] allows determining the energies of pure spin states quite correctly.

Answer 2 concerns the quantitative description of graphene magnetism and shows that broken symmetry approaches provide exact determination of magnetic constant. The value is size-dependent and steadily increases when the graphene molecule size increases. The molecules with linear dimension of a few *nm* can provide the constant small enough for the magnetism of singlet graphene to be recorded. However, when the size exceeds the electron mean free pass, the magnetism disappears due to quantizing electronic states and coming back to crystalline graphene unit cell that is diamagnetic.

Answer 3 is related to the graphene characteristic that controls the odd electrons correlation. As shown, this is the C-C bond length that exceeds the critical value $R_{crit}$ =1.395 Å. Above this value two adjacent odd electrons become effectively unpaired, firstly, partially radicalized and then completely radicalized as the C-C distance grows.

Answer 4 addresses a definite physical reality of effectively unpaired electrons. So far there has been only one case when UBS HF computational results were compared with those obtained by using one of the CI schemes in the form of either CASSCF or MRCI approach [38]. The two techniques were applied to the description of diradical character of the Cope rearrangement transition state. CASSCF, MRCI, and UBS HF calculations have revealed effectively unpaired electrons $N_D$ at the level of 1.05, 1.55, and 1.45 *e*, respectively, just highlighting that the feature is a characteristic for the electron correlation but not the proximity of the UBS HF approach. Recent successes in developing atomic force microscopy with unprecedented high accuracy have allowed seeing the unpaired electrons directly. The recorded molecular images for pentacene, olympicene, and graphene molecules are in full consent with those calculated in the UBS HF approximation.

Answer 5 concerns the basic grounds of chemical modification of graphene. As shown, the fractional number of effectively unpaired electrons related to a given atom $N_{DA}$ is a quantitative indication of the atom chemical activity (atomic chemical accessibility) that can be used as a reliable indicator of the target atom entering the reaction. A large scale stepwise reaction can be considered computationally, which leads to formation of different polyderivatives of graphene. An example of hydrogenation and oxidation of graphene, was obtained a general view of graphene polyhydrides and polyoxides that well fit experimental reality.

Answer 6 testifies that molecular theory is quite efficient when considering mechanical behavior of graphene. Leaving outside the theory application to the consideration of the deformational process as such [85-87], the current paper is concentrated on the consequences, related to odd electros correlation, that are caused by changing C-C bond lengths in graphene in due course of deformation. The deformation might be either dynamic or static. The former is caused by an external loading of a graphene sample while the latter concerns the deformation of the carbon skeleton of graphene due to chemical modification. In both cases, stretching C-C bond causes increasing of both the total and fractional numbers of effectively unpaired electrons. The feature explains changing in chemical reactivity of graphene during deformation, on one hand, and appearing bright spots on TEM images in the area of graphene bubbles.

A limited volume of the paper does not allow touching all features of extremely large graphene science. However, the selected topics and answers obtained in due course of their consideration clearly show that molecular theory of graphene, implemented in the format of UBS HF computing schemes, is highly efficient and suggests reliable explanations for a number of different graphene peculiarities. These explanations are obtained on the same platform based on quite a few concepts involving odd electrons of the graphene benzenoid units and their correlation due to weak interaction. Outside the paper, there are still questions concerning chemical topology of graphene [86, 87], different aspects of intermolecular interaction with impurity [88], graphene-base shungite quantum dots formation [89], graphene catalytic activity [90, 91], silicene as siliceous counterpart of graphene [92, 93], and so forth. And molecular theory of graphene is very successful in dealing with all these issues, not being concentrated on numbers but giving the main attention to clearly seen trends.

The odd electron correlation is not a prerogative of graphene only. Similar phenomenon is characteristic for all $sp^2$ nanocarbons, including both fullerenes and nanotubes [5]. The only preference of graphene consists in much larger variety of cases when this inherent characteristic of the class can be visualized.

**Acknowledgments**

The author immensely appreciates fruitful discussions with I.L.Kaplan, E.Brandas, D.Tomanek. O.Ori, F. Cataldo, E.Molinary L.A.Chernozatonski who draw her attention onto different problems of the molecular theory of graphene. The author is deeply grateful to her colleagues N.Popova, V.Popova, L. Shaymardanova, B.Razbirin, D. Nelson, A. Starukhin, N.Rozhkova for support and valuable contribution into the study.

REFERENCES

1. Hoffmann R (2013) Small but strong lessons from chemistry for nanoscience. Ang Chem Int Ed 52: 93-103
2. Hoffmann R (1971) Interaction of orbitals through space and through bonds. Acc Chem Res 4 : 1–9
3. Hay PJ, Thibeault JC, Hoffmann R (1971) Orbital interactions in metal dimer complexes. J Amer Chem Soc 97: 4884-4899
4. Sheka E (2003) Violation of covalent bonding in fullerenes. In: Sloot P M A, Abramson D, Bogdanov A V et al (eds) Lecture Notes in Computer Science, Computational Science – ICCS2003, Springer, Heidelberg, p 386-398
5. Sheka EF (2011) Fullerenes: Nanochemistry, nanomagnetism, nanomedicine, nanophotonics. CRC Press, Taylor and Francis Group, Boca Raton
6. Sheka EF (2003) Fullerenes as polyradicals. Internet Electronic Conference of Molecular Design, 2003, 23 November – 6 December 2003. http://www.biochempress.com, November 28, paper 54
7. Sheka EF (2004) Odd electrons and covalent bonding in fullerenes. Int J Quant Chem 100: 375-386
8. Sheka E. (2009) Nanocarbons through computations: Fullerenes, nanotubes, and graphene. In: The UNESCO-EOLSS Encyclopedia Nanoscience and Nanotechnology. UNESCO, Moscow, p. 415-444
9. Geim AK, Novoselov KS (2007) The rise of graphene. Nature Mat 6: 183-191
10. Davidson E (1998) How robust is present-day DFT? Int J Quant Chem 69: 214-245.
11. Kaplan I (2007) Problems in DFT with the total spin and degenerate states. Int J Quant Chem 107: 2595-603
12. Takatsuka K, Fueno T, Yamaguchi K (1978) Distribution of odd electrons in ground-state molecules. Theor Chim Acta 48: 175-83
13. Staroverov VN, Davidson E R (2000) Distribution of effectively unpaired electrons. Chem Phys Lett 330: 161-168
14. Benard M J (1979) A study of Hartree–Fock instabilities in $Cr_2(O_2CH)_4$ and $Mo_2(O_2CH)_4$. J Chem Phys 71: 2546-56
15. Lain L, Torre A, Alcoba D R et al (2011) A study of the relationships between unpaired electron density, spin-density and cumulant matrices. Theor Chem Acc 128: 405-410

16. Sheka EF, Chernozatonskii LA (2007) Bond length effect on odd electrons behavior in single-walled carbon nanotubes. J Phys Chem A 111: 10771-10780
17. Sheka E F (2012) Computational strategy for graphene: Insight from odd electrons correlation. Int J Quant Chem 112: 3076-3090
18. Zayets VA (1990) CLUSTER-Z1: Quantum-chemical software for calculations in the *s,p*-basis. Institute of Surface Chemistry Nat Ac Sci of Ukraine: Kiev
19. Gao X, Zhou Z, Zhao Y et al (2008) Comparative study of carbon and BN nanographenes: Ground electronic states and energy gap engineering. J Phys Chem A 112: 12677-82
20. Noodleman L (1981) Valence bond description of antiferromagnetic coupling in transition metal dimers. J Chem Phys 74: 5737-42
21. Illas F, Moreira I de P R, de Graaf C, Barone V (2000) Magnetic coupling in biradicals, binuclear complexes and wide-gap insulators; a survey of *ab initio* function and density functional theory approaches. Theor Chem Acc 104: 265-272
22. Zvezdin AK, Matveev VM, Mukhin AA et al (1985) Redkozemeljnyje iony v magnito-uporjadochennykh kristallakh (Rear Earth ions in magnetically ordered crystals) Nauka, Moskva
23. Van Fleck JH (1932) The theory of electric and magnetic susceptibilities. Oxford at the Clarendon Press, Oxford
24. Kahn O (1993) Molecular Magnetism. VCH, New York
25. Koshino M, Ando T (2007) Diamagnetism in disordered graphene. Phys Rev B 75: 235333 (8 pages)
26. Nair RR, Sepioni M, Tsai I-L et al (2012) Spin-half paramagnetism in graphene induced by point defects. Nature Phys 8: 199-202
27. Sheka EF, Chernozatonskii LA (2010) Chemical reactivity and magnetism of graphene. Int J Quant Chem 110: 1938-1946
28. Sheka EF, Chernozatonskii LA (2010) Broken spin symmetry approach to chemical susceptibility and magnetism of graphenium species. J Exp Theor Phys 110: 121-132
29. Shibayama, Y.; Sato, H.; Enoki, T.; Endo, M. Phys. Rev. Lett. 2000, 84, 1744.
30. Enoki, T.; Kobayashi, Y. J. Mat. Chem. 2005, 15, 3999.
31. Tada K, Haruyama J, Yang H X et al (2011) Graphene magnet realized by hydrogenated graphene nanopore arrays. Appl Phys Lett 99: 183111(3 pages)
32. Tada K, Haruyama J, Yang H X et al (2011) Ferromagnetism in hydrogenated graphene nanopore arrays. Phys Rev Lett 107: 217203 (5 pages)
33. Sheka EF, Zayets VA, Ginzburg IYa (2006) Nanostructural magnetism of polymeric fullerene crystals. J Exp Theor Phys 103: 728-739
34. Boeker GF (1933) The diamagnetism of carbon tetrachloride, benzene and toluene at different temperatures. Phys Rev 43: 756-760
35. Seach MP, Dench WA (1979) Quantitative electron spectroscopy of surfaces: A standard data base for electron inelastic mean free paths in solids. Surf Interf Anal 1: 2-11
36. Komolov SA, Lazneva EF, Komolov AS (2003) Low-energy electron mean free path in thin films of copper phthalocyanine. Tech Phys Lett 29: 974-976

37. Takatsuka K, Fueno T J (1978) The spin-optimized SCF general spin orbitals. II. The $2\,^2S$ and $2\,^2P$ states of the lithium atom. J Chem Phys 69:661-669
38. Staroverov VN, Davidson ER (2000) Diradical character of the Cope rearrangement transition state. J Am Chem Soc 122: 186-187
39. Mayer I (1986) On bond orders and valences in the *ab initio* quantum chemical theory. Int J Quant Chem 29: 73-84
40. Dewar MJS, Thiel W (1977) Ground states of molecules. 38. The MNDO method. Approximations and parameters. J Am Chem Soc 99: 4899-4907
41. Zhogolev DA, Volkov VB (1976) Metody, algoritmy i programmy dlja kvantovo-khimicheskikh raschetov molekul (Methods, algorithms and programs for quantum-chemical cCalculations of molecules) Kiev, Naukova Dumka
42. Sheka EF, Zayets VA (2005) The radical nature of fullerene and its chemical activity. Russ J Phys Chem 79: 2009-2014
43. Lain L, Torre A, Alcoba DR et al (2009) A decomposition of the number of effectively unpaired electrons and its physical meaning. Chem Phys Lett 476: 101-103
44. Wang J, Becke AD, Smith VHJr (1995) Eveluation of $\langle \hat{S}^2 \rangle$ in restricted, unrestricted Hartree-Fock, and density functional based theory. J Chem Phys 102: 3477-3480
45. Cohen AJ, Tozer DJ, Handy N C (2007) Evaluation of $\langle \hat{S}^2 \rangle$ in density functional theory. J Chem Phys 126: 214104 (4pp)
46. Lobayan RM, Bochicchio RC, Torre A et al (2011) Electronic structure and effectively unpaired electron density topology in *closo*-boranes: Nonclassical three-center two-electron bonding. J Chem. Theory Comp 7: 979-987
47. Kitagawa Y, Saito T, Ito M et al (2007) Approximately spin-projected geometry optimization method and its application to di-chromium systems. Chem Phys Lett 442: 445-450
48. Kitagawa Y, Saito T, Nakanishi Y et al (2009) Spin Contamination Error in Optimized Geometry of Singlet Carbene ($^1A_1$) by Broken-Symmetry Method. J Phys Chem A 113: 15041-15046
49. Gross L, Mohn F, Moll N et al (2009) The chemical structure of a molecule resolved by atomic force microscopy. Science 325:1110-1114
50. 'Olympic rings' molecule olympicene in striking image BBC News Science and Environment (2012-05-28)
51. Fujita M, Wakabayashi K, Nakada K et al (1996) Peculiar localized state at zigzag graphite edge. J Phys Soc Jpn 65: 1920-1923
52. Nakada K, Fujita M, Dresselhaus G et al (1996) Edge state in graphene ribbons: Nanometer size effect and edge shape dependence. Phys Rev B 54: 17954-17961
53. Coleman J (2008) A new solution to graphene production. SPIE Newsroom. DOI: 10.1117/2.1200810.1336
54. The noise about graphene (2010) Science Centre of Barkley Lab

55. Sheka EF (2006) 'Chemical portrait' of fullerene molecule. J Str Chem 47:, 600-607
56. Sheka EF (2007) Chemical susceptibility of fullerenes in view of Hartree-Fock approach. Int Journ Quant Chem 107: 2803-2816
57. Sheka EF, Chernozatonskii LA (2010) Chemical reactivity and magnetism of graphene. Int J Quant Chem 110: 1938-1946
58. Sheka EF, Chernozatonskii LA (2010) Broken spin symmetry approach to chemical susceptibility and magnetism of graphenium species. J Expt Theor Phys 110: 121-132
59. Allouche A, Jelea A, Marinelli F et al (2006) Hydrogenation and dehydrogenation of graphite (0001) surface: a density functional theory study. Phys. Scr. T124: 91-94
60. Sheka EF (2010) Stepwise computational synthesis of fullerene $C_{60}$ derivatives. Fluorinated fullerenes $C_{60}F_{2k}$. J Expt Theor Phys 111: 395-412
61. Sheka EF, Popova NA (2012) Odd-electron molecular theory of the graphene hydrogenation. J Mol Mod 18: 3751-3768
62. Elias DC, Nair RR, Mohiuddin TMG et al (2009) Control of Graphene's Properties by Reversible Hydrogenation: Evidence for Graphane. Science 323: 610-613
63. Sheka EF (2011) Computational synthesis of hydrogenated fullerenes from $C_{60}$ to $C_{60}H_{60}$. J Mol Mod 17:1973-1984
64. Sheka EF, Popova NA (2011) When a covalent bond is broken? arXiv:1111.1530v1 [physics.chem-ph]
65. Sheka EF, Popova NA (2012) Molecular theory of graphene oxide. arXiv:1212.6413 [cond-mat.mtrl-sci]
66. Sheka EF, Popova NA (2012) Molecular theory of graphene oxide. Phys Chem Chem Phys (delivered)
67. Dreyer DS, Park S, Bielawski CW et al (2010) The chemistry of graphene oxide. Chem Soc Rev 39: 228–240
68. Zhu Y, Shanthi M, Weiwei C et al (2010) Graphene and Graphene Oxide: Synthesis, Properties, and Applications. Adv Mater 22: 3906–3924
69. Kuila T, Bose S, Mishra AK et al (2012) Recent advances in the efficient reduction of graphene oxide and its application as energy storage electrode materials. Progr Mat Sci 57: 1061–1105
70. Wang H, Hu IH (2011) Effect of Oxygen Content on Structures of Graphite Oxides. Ind Eng Chem Res 50: 6132–6137
71. Fujii S, Enoki T (2010) Cutting of Oxidized Graphene into Nanosized Pieces. J Am Chem Soc 132: 10034–10041
72. Xu Z, Bando Y, Liu L et al (2011) Electrical conductivity, chemistry, and bonding alternations under graphene oxide to graphene transition as revealed by *in situ* TEM. ACS Nano 5: 4401–4406.
73. Wang S, Wang R, Liu X et al (2012) Optical Spectroscopy Investigation of the Structural and Electrical Evolution of Controllably Oxidized Graphene by a Solution Method. J Phys Chem C 116: 10702-10707

74. Mattevi C, Eda G, Agnoli S et al (2009) Evolution of electrical, chemical, and structural properties of transparent and conducting chemically derived graphene thin films. Adv Funct Mat 19: 2577-2583
75. Sheka EF (2010) Computational synthesis of hydrogenated fullerenes from $C_{60}$ to $C_{60}H_{60}$. J Mol Mod 17: 1973-1984.
76. Wang L, Zhao J, Sun Y -Y et al (2011) Characteristics of Raman spectra for graphene oxide from *ab initio* simulations. J Chem Phys 135: 184503 (5 pages)
77. Saxena S, Tyson TA, Negusse E (2010) Investigation of the Local Structure of Graphene Oxide. J Phys Chem Lett 1: 3433–3437
78. Ambrosi A, Chee SY, Khezri B et al (2012) Metallic impurities in graphenes prepared from graphite can dramatically influence their properties. Angew Chem Int Ed 51: 500 –503
79. Lu N, Li Zh (2012) Graphene oxide: Theoretical perspectives. In J. Zeng et al. (eds.), Quantum Simulations of Materials and Biological Systems, Springer Science+Business Media Dordrecht, pp69-84.
80. Levy N, Burke S A, Meaker KL et all (2010) Strain-Induced Pseudo–Magnetic Fields Greater Than 300 Tesla in Graphene Nanobubbles Science 329: 544-547
81. Georgiou T, Britnell L, Blake P et al (2011) Graphene bubbles with controllable curvature. Appl Phys Lett **99:** 093103 (3 pages)
82. Koenig SP, Boddeti NG, Dunn ML et al (2011) Ultrastrong adhesion of graphene membranes. Nature Nanotechn. 6: 543-546
83. Sheka EF, Popova NA, Popova VA et al (2011) Structure-sensitive mechanism of nanographene failure. J Exp Theor. Phys 112: 602-611
84. Sheka EF, Popova NA, Popova VA et al (2011) A tricotage-like failure of nanographene. J Mol Mod 17: 1121-1131
85. Popova NA, Sheka EF (2011) Mechanochemical reaction in graphane under uniaxial tension. J Phys Chem C 115 23745-23754
86. Sheka EF, Shaymardanova LKh (2011) $C_{60}$-based composites in view of topochemical reactions. J Mater Chem 21:17128 – 17146
87. Sheka EF (2013) Topochemistry of spatially extended *$sp^2$* nanocarbons: fullerenes, nanotubes, and graphene. In Ashrafi A R, Cataldo F, Iranmanesh A et al (Eds.) Topological Modelling of Nanostructures and Extended Systems. Carbon Materials: Chemistry and Physics, vol. 7, Springer, Heidelberg, pp. xxx-yyy
88. Sheka EF, Razbirin BS, Rozhkova NN et al (2012) Nanophotonics of graphene quantum dots. Paper presented at the XV International Conference "Laser Optics-2012" St.Petersburg, Russia, 25-29 June 2012
89. Sheka EF, Rozhkova NN (2013) New carbon allotrope shungite as loosely packed fractal nets of graphene-base quantum dots. To be published
90. Tang S, Cao Z (2012) Site-dependent catalytic activity of graphene oxides towards oxidative dehydrogenation of propane. Phys Chem Chem Phys 14: 16558-16565
91. Hsu H-C, Shown I, Wei H-Y et al (2013) Graphene oxide as a promising photocatalyst for $CO_2$ to methanol conversion. Nanoscale 5: 262-268
92. Sheka EF (2009) May silicene exist? http://arXiv.org/abs/0901.3663

93. Sheka EF (2013) Why sp2-like nanosilicons should not form: Insight from quantum chemistry. Int J Quant Chem 113: 612-618